\documentclass[final,1p,times]{elsarticle}

\usepackage{amssymb}

\usepackage{amsmath}
\usepackage{caption}
\usepackage{subcaption}

\journal{Acta Materialia}

\begin{document}

\begin{frontmatter}



\title{Feature engineering for microstructure-property mapping in organic photovoltaics }


\author[inst1]{Sepideh Hashemi}
\affiliation[inst1]{organization={George W. Woodruff School of Mechanical Engineering, Georgia Institute of Technology},
            addressline={}, 
            city={Atlanta},
            postcode={30332}, 
            state={GA},
            country={USA}}

\author[inst2]{Baskar Ganapathysubramanian}
\affiliation[inst2]{organization={Department of Mechanical Engineering, Iowa State University},
            addressline={}, 
            city={Ames},
            postcode={50011}, 
            state={IA},
            country={USA}}

\author[inst3]{Stephen Casey}            
\author[inst3]{Ji Su}
\affiliation[inst3]{organization={NASA Langley Research Center},
            addressline={}, 
            city={Hampton},
            postcode={23681}, 
            state={VA},
            country={USA}}

\author[inst1,inst4]{Surya R. Kalidindi\corref{cor}}
\affiliation[inst4]{organization={School of Computational Science and Engineering, Georgia Institute of Technology},
            addressline={}, 
            city={Atlanta},
            postcode={30332}, 
            state={GA},
            country={USA}}
\cortext[cor]{Corresponding author at: George W. Woodruff School of Mechanical Engineering, Georgia Institute of Technology, Atlanta, GA 30332, USA. E-mail address: surya.kalidindi@me.gatech.edu (S.R. Kalidindi).}

\begin{abstract}
Linking the highly complex morphology of organic  photovoltaic (OPV) thin films to their charge transport properties is critical for achieving high performance material system that serves as a cost-efficient approach for energy harvesting. In this paper, a novel unsupervised feature engineering framework is developed and used to establish reduced-order structure-property linkages for OPV films. This framework takes advantage of digital image processing algorithms to identify the salient material features of OPVs undergoing the charge transport phenomenon. These material states are then used to obtain a low-dimensional representation of OPV microstructures via 2-point spatial correlations and principal component analysis. It is found that in addition to the material PC scores, two distance-based metrics are required to complete the microstructure quantification of complex OPVs. A localized version of the Gaussian process (laGP) is then used to link the material PC scores as well as the two distance-based metrics to the short-circuit current of OPVs. It is demonstrated that the unsupervised feature engineering framework presented in this paper in conjunction with the laGP can lead to high-fidelity and accurate data-driven structure-property linkages for OPV films.
\end{abstract}




\end{frontmatter}




\section{Introduction}
\label{sec:Intro}
	
Flexible, light-weight, and wearable solar cells offer a promising solution to cheap energy harvesting for consumer products as well as residential applications.  Over the past decade, rapid developments in synthetic chemistry have resulted in organic photovoltaic systems that have pushed single-junction organic photovoltaic (OPV) efficiencies over 16\%. These novel materials -- electron-donors and electron-acceptors -- provide tremendous opportunities for improved performance, reaching the performance of silicon based photovoltaics. In conjunction with synthesis advances, a large body of work has demonstrated that the microstructure in the active layer is key to high performance devices. Thus, tailoring the morphology in the active layer of OPVs continues to be crucial for maximizing performance. More importantly, advances in self-assembly suggests the possibility of remarkable control of the active layer morphology. 

Despite the importance of morphology to OPV device performance, it remains a challenge to comprehensively and rapidly map morphologies to performance. The availability of reliable and fast structure-property models could enable domain scientists to (a) explore, identify and design "ideal" morphologies that maximize performance, (b) identify microstructure features that positively (or negatively) impact performance, and (c) quantify how perturbations to the morphology (due to oxidation, annealing or ageing) degrade performance. 

Past approaches of investigating structure-property linkages relied on full-physics simulators --- either discrete (kinetic) monte carlo models, or continuum drift-diffusion models. These models are typically expensive to deploy, and sequential deployment for exploration or optimization has been shown to be prohibitively expensive. Similarly, rapid design exploration using such full-physics simulators is typically not possible, even with access to high performance computing resources.

Recent approaches overcome this challenge by first creating a diverse dataset of annotated morphologies and their performances, and then utilizing data-driven tools on this dataset to construct \textit{low-computational cost surrogate structure-property models}. Such a strategy amortizes the cost of creating a large annotated dataset across multiple studies. Additionally, property annotation on this dataset using the full-physics simulators are embarrassingly parallel, thus, optimally utilizing HPC resources. 

Such structure-property surrogate models -- especially in the context of OPV -- have been successfully constructed and deployed for design optimization, process-structure-property linkages, sensitivity analysis, and other studies. However, most of these studies have: 
\begin{itemize}
\item either relied on manual 'featurization' of the morphologies based on knowledge of the photophysics~\citep{wodo2013quantifying,wodo2012computational,wodo2015automated}. While very useful, such approaches are non-trivial and generally time-consuming. Additionally, manual featurization carries the risk of overlooking or neglecting important features,
\item or utilized the full raw morphology data to construct structure-property linkages~\citep{pokuri2019interpretable}. However, these approaches need massive datasets to train good surrogate models due to the large input dimensionality (of the morphology image). Additionally, the resultant surrogates are complex and usually not interpretable. 
\end{itemize}

In this work, we bridge these two extremes by using a principled approach of unsupervised featurization of the morphologies. These low dimensional set of features are then used to train an accurate structure-property surrogate model. Specifically, the recently developed Material Knowledge System (MKS) framework \citep{Kalidindi2015,Iskakov2018,latypov2019,Yabansu2020,Hashemi2021} offers a data-driven framework for unsupervised feature engineering of material microstructures. This framework employs a voxelized representation of microstructures to efficiently compute the 2-point spatial correlations \citep{Torquato2002, Niezgoda2011,Niezgoda2013} and perform principal component analysis (PCA) \citep{Hastie2005,Bishop2006} on them to identify a sufficiently small number of features representing the complex material microstructure. The feature engineering developed in the MKS framework is unsupervised in that the microstructure feature selection is completely uninfluenced by the output variables targeted by the surrogate model. Although a large number of options exist for building the surrogate models of interest, recent work in the MKS framework \citep{Yabansu2020,Hashemi2021,Fernandez2019,Tallman2019,Yabansu2019,Yabansu2019GPAR,Parvinian2020,Marshal2021}  has demonstrated that Gaussian process regression (GPR) \citep{Bishop2006, Williams2006} offers advantages because of its ability to formulate non-parametric models while allowing for a rigorous consideration of the prediction uncertainty. 

Certain extensions are needed to the current MKS framework in order to  apply it successfully to the present problem. First, a large number of pixel-scale (local) material states need to be considered, which is expected to be significantly larger than those encountered in prior case studies. This is because of the need to consider not only the donor and acceptor pixels, but also the different types of the donor-acceptor interfaces present in the microstructure. Second, the small thickness of the films requires additional considerations in the feature engineering. This is because the distances of the different types of the donor-acceptor interfaces from the top and the bottom surfaces of the films are known to control their effective properties \citep{Wodo2012}.

This paper describes the extensions made to the MKS framework so that it can be applied successfully to establish data-driven microstructure-property linkages for OPV films. More specifically, this paper develops and demonstrates novel approaches to feature engineering the complex microstructures in blend films by combining digital image processing techniques with the previously established MKS feature selection methods. The employment of digital image processing techniques allows for computationally efficient pixel-scale labeling of the different material states in the polymer blend films. When these protocols are followed by 2-point spatial correlations and PCA, they offer novel unsupervised feature engineering of the complex material structure of OPV films. The tremendous utility of such feature engineering protocols is demonstrated in this paper by building a surrogate model for the prediction of the short circuit current of photovoltaic polymer blend films using a localized version of GPR.

\section{Background}

\subsection{Microstructure and photovoltaic property dataset}
\label{sec:data_gen}

We utilize a curated dataset of microstructure images created by solving the Cahn-Hilliard equation~\citep{cahn1958free} with varying initial conditions. The Cahn-Hilliard equation~\citep{cahn1958free} describes phase separation occurring in a binary mixture, and has been shown to be a good representation of morphology evolution during fabrication of organic blend thin films~\citep{wodo2012modeling,wodo2014evaporating,zhao2016vertical} that are the typical active layer in OPV's. The image data arising from these simulations provide a rich dataset for constructing structure-property surrogate models~\citep{pokuri2019interpretable}. The dataset is a collection of $33,552$ microstructure images of $101 \times 101$ pixels in resolution. Each image is grayscale, with the value of each pixel ranging between $0$ to $1$. 

Each microstructure is virtually interrogated to extract its current-voltage characteristics, by solving a morphology aware (i.e., spatially heterogeneous) photophysics device model. We deploy a validated, in-house software that uses a finite element based solution strategy for solving the photophysics device model~\citep{kodali2012computer,kodali2012computational,pfeifer2018process}. The photophysics model is described by the steady state \textit{excitonic drift diffusion (XDD) equations}. The XDD equations are a set of four tightly coupled partial differential equations that model the optoelectronic physics of energy harvesting in organic photovoltaic devices. The photophysics
consists of the following stages (also illustrated in Figure \ref{fig:XDD}): 

\begin{figure}
    \centering
    \includegraphics[width=0.9\textwidth]{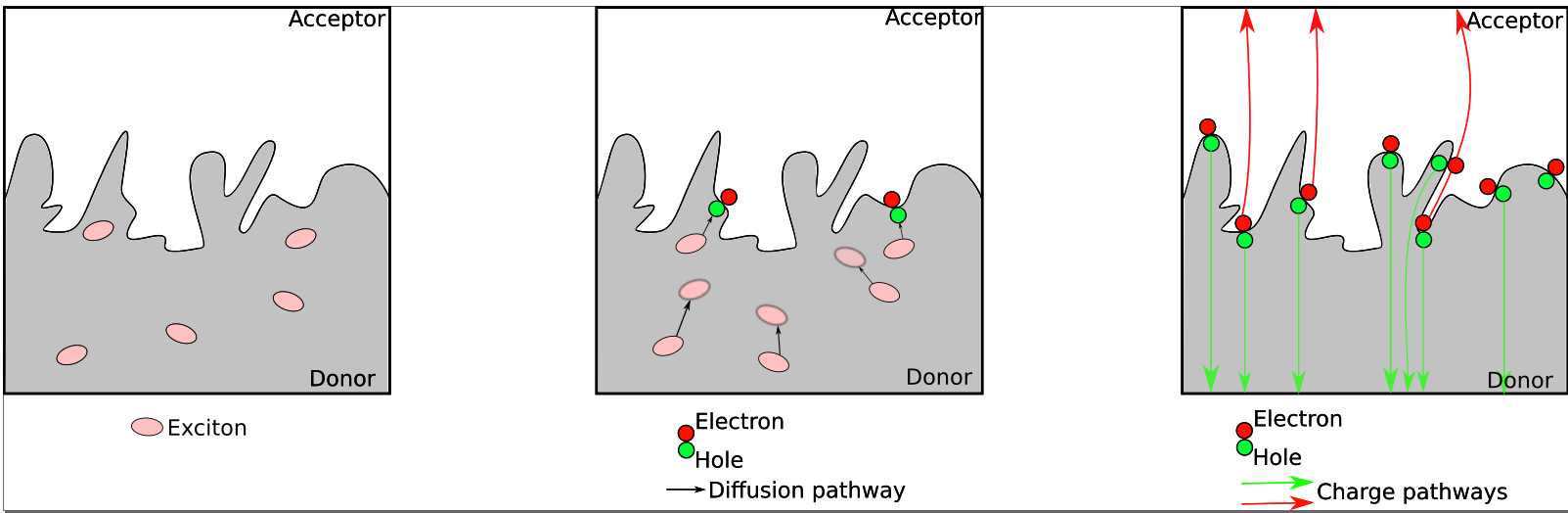}
    \caption{Schematic illustrating the various stages of the photophysics process (see main text for detailed description).}
    \label{fig:XDD}
\end{figure}
\begin{itemize}
    \item Incident solar radiation causes the generation of energetically active electron-hole pairs, called excitons (denoted by X), in the donor regions of the microstructure. These excitons diffuse across the microstructure and have a finite lifetime before becoming ground state electron-hole pairs;
    \item Excitons that diffuse and reach the donor-acceptor interface undergo dissociation into electrons (denoted by n) and holes (denoted by p) at the donor-acceptor interface. The dissociation mechanism is material and field dependent (denoted by D); 
    \item These generated charges (n,p) traverse the microstructure and reach their corresponding electrodes (cathode and anode) to produce a current. Two mechanisms are responsible for driving carrier transport or current flow. First, the drift, which is caused by the presence of an electric field (denoted as the gradient of the potential, $\nabla \varphi$, and second, the diffusion, which is caused by a spatial gradient of electron or hole concentration;
    \item  The distribution of electrons and holes in the microstructure interacts with the applied voltage and influences the electrostatic potential $\varphi$ across the microstructure. Finally, electrons and holes can recombine (denoted by R) to create excitons
\end{itemize}

The photophysics described above is encoded using the exciton drift diffusion (XDD) equations~\citep{kodali2012computer}. 
In prior work, these XDD equations were solved to get the performance of the OPV device, which is characterized by the short-circuit current $J_{sc}$. XDD simulation results for each of the 34672 microstructures generated earlier provide us the photophysics properties ($J_{sc}$).

\subsection{Feature engineering using MKS framework}
\label{sec:MKS_featureENG}

In the MKS framework, the uniformly discretized (i.e., voxelated) representative volume elements (RVEs) of the material microstructures are denoted by an array, $m_{\boldsymbol s}^{h}$, whose elements denote the volume fractions of the material state $h$ found at voxel $\boldsymbol s$. Microstructural domains where each voxel is occupied fully by a specific material state leads to microstructure arrays where the value of $m_{\boldsymbol s}^{h}$ is either $0$ or $1$. Although it may be tempting to use $m_{\boldsymbol s}^{h}$  directly as the feature set, it should be recognized that it lacks translational invariance. The MKS framework employs the framework of 2-point spatial correlations \citep{Torquato2002, Niezgoda2011, Niezgoda2013}, which are essentially auto- and cross-correlations of material state maps of the microstructure. Mathematically, the discretized set of 2-point spatial correlations, denoted as $f_{\boldsymbol r}^{h h^{'}}$, are computed as
\begin{equation} \label{2-pt}
    f_{\boldsymbol r}^{h h^{'}} = \frac{1}{S_{\boldsymbol r}} \sum_{\boldsymbol s} m_{\boldsymbol s}^{h} m_{\boldsymbol {s+r}}^{h^{'}}
\end{equation}
\noindent where $h$ and $h^{'}$ index all of the material states present in the studied material system, $\boldsymbol r$ indexes a set of discretized vectors arising from the voxelization used to define $m_{\boldsymbol s}^{h}$, and $S_{\boldsymbol{r}}$ denotes the total number of pixels that allow for placement of vectors ${\boldsymbol{r}}$ within the microstructural domain. The computations implied in Eq. (1) can be efficiently carried out using the fast Fourier transform (FFT) algorithm \citep{Fullwood2010, Cecen2016}.

The complete set of 2-point spatial correlations computed using Eq.\eqref{2-pt} produces a large unwieldy set of features. In the MKS framework, a smaller set of salient features is identified (i.e., feature engineering) by performing principal component analysis (PCA) \citep{Hastie2005,Bishop2006}, which (rotationally) transforms the data into a new space where the axes are organized by their ability to account for the variance in the dataset. The new orthogonal axes and the new coordinates obtained from the PCA are then referred to as PC scores and PC basis, respectively. Prior studies have often shown a drastic dimensionality reduction going from $\sim \! 10^5 - 10^6$ original microstructural features to less than $ \sim \! 10-15$ PCs \citep{Iskakov2018, Yabansu2020, Khosravani2017, Paulson2017, Fernandez2019}.

\subsection{Gaussian process regression models }

Although many surrogate model building approaches can be used for building structure-property linkages, prior work has shown the benefits of using Gaussian process regression (GPR) in combination with the MKS feature engineering described earlier \citep{Yabansu2020,Hashemi2021,Fernandez2019,Tallman2019,Yabansu2019,Yabansu2019GPAR,Parvinian2020,Marshal2021}. GPR is particularly powerful when building surrogate models for complex nonlinear systems/phenomena, where the parametric model forms are not yet established. The other main advantage of GPR lies in the quantification of the uncertainty associated with the model predictions.
 
In the GPR-MKS framework, the reduced-order structure-property linkage of interest can be decomposed into a linear mean function $m$  and  an error function $\varepsilon$ often modeled as a zero-mean Gaussian process. Mathematically, the desired model is expressed as \citep{Williams2006}
\begin{align} \label{S-P}
   & p = m(\boldsymbol \gamma) + \varepsilon \\
        & m(\boldsymbol \gamma) =  \beta_0 + \sum_{i=1}^{R} \beta_i \gamma_i \\
    & \varepsilon \sim  \mathcal{GP}(0,k(\boldsymbol \gamma, \boldsymbol {\gamma^{'}})) \label{eqn:GP}
\end{align}
\noindent where $p$ is the  target property (i.e., output), $\boldsymbol \gamma$ is the input feature vector consisting of $R$ PCs, $\boldsymbol \beta$ are coefficients of the linear model, and  $k(\boldsymbol \gamma, \boldsymbol {\gamma^{'}})$ is the GP's covariance function. The automatic relevance determination squared exponential (ARD-SE) kernel \citep{Williams2006} has often been used to define the GP's covariance. The ARD-SE kernel is mathematically expressed as

\begin{equation}
    k(\boldsymbol{\gamma , \gamma^{'}})=\sigma_{f}^{2} \exp{\left[ -\frac{1}{2} \sum_{l=1}^{R} \frac{\left( \gamma_l - \gamma_{l}^{'} \right) ^2}{\sigma_{l}^{2}} \right]} + \sigma_{n}^{2} \delta_{\boldsymbol{\gamma \gamma^{'}}}
\end{equation}
where the scaling factor $\sigma_f$, length scale $\sigma_l$, and noise factor $\sigma_n$ are hyperparameters of the kernel function, and $\delta_{\boldsymbol{\gamma \gamma^{'}}}$ is the Kronecker delta. The hyperparameter $\sigma_n$ determines the homoscedastic noise in the target predictions. The hyper parameter $\sigma_f$ controls the amplitude of the variance in the output. The length scale $\sigma_l$ automatically determines the relevance of input features on the predictions. Higher values of $\sigma_l$ results in smoother predictions, indicating minimal influence on the output prediction. The values of hyperparameters need to be optimized during the model building process to obtain the best model.

The joint distribution of the observed training data ($\boldsymbol X$) and the unobserved test data ($\boldsymbol X_*$) is given by \citep{Williams2006}

\begin{equation}
  \left[ \begin{array}{c} \boldsymbol p \\ \boldsymbol p_*  \end{array} \right] \sim \mathcal{N} \left( \boldsymbol 0 \; , \left[ \begin{array}{cc} K(\boldsymbol{X,X}) & K_*(\boldsymbol{X,X_*}) \\
K_*^\dagger(\boldsymbol{X,X_*}) & K_{**}(\boldsymbol{X_*,X_*}) \end{array} \right] \right)
\end{equation}

\noindent The predictive posterior is obtained from conditioning the joint distribution fully defined by its mean and covariance \citep{Williams2006}:

\begin{equation}
\begin{split}
     \boldsymbol{\mu_*} &= K_{*}^{\dagger} K^{-1} \boldsymbol p \\
     \boldsymbol{\Sigma_*} &= K_{**} - K_{*}^{\dagger} K^{-1} K_*
\end{split}
\end{equation}

\noindent The main computationally intensive operation in GP formulation is the inversion of the kernel matrix which scales as $O(N^3)$. Although this is a one-time computation, in case of large ensemble of training data, the computation and storage of $K^{-1}$ present significant challenges. Prior studies have addressed these challenges using methods such as low-rank approximations to GPs \citep{Williams2006,Wilson2015}, treed GPs \citep{Bui2014,Lee2017} and local approximate GP (laGP) \citep{Gramacy2015,Gramacy2016}. Recent research has demonstrated that low-rank approximations and treed GPs tend to over-smooth the data, might impose an upper limit on the data size and typically take longer to compute \citep{Heaton2019}. The recently developed laGP model is particularly attractive as it scales well with the data size, allows for non-stationarity modelling, and is highly parallelizable. The laGP model employs a local subset of the data to train separate GPs for each target point. The subset of data can be chosen as $n$ nearest neighbors of the target point. However, this simple criterion does not yield the optimum predictions. Instead, the laGP approach utilized in this work employs the active learning Cohn (ALC) method \citep{Gramacy2016,Cohn1996} to sequentially update the chosen subset of the training points. The ALC sequentially identifies points whose addition to the local subset maximizes the expected information gain by maximizing the reduction in the prediction variance.

\section{Microstructure-Property models for photovoltaic polymers}
\label{sec:S-P}

The workflow used in this paper for building the surrogate microstructure-property models for OPVs will involve two main steps: (i) unsupervised feature engineering of the microstructure using the MKS framework, and (ii) establishing the laGP models using the engineered features. Further details of these steps are described next.

\subsection{Material states in OPV microstructures}
\label{sec:feature_eng}

The gray-scale OPV  microstructures (with each pixel value ranging between zero and one) obtained from solving the Cahn-Hilliard equation (summarized in section \ref{sec:data_gen}) are thresholded into binary microstructures consisting of donor ($D$) and acceptor ($A$) phases (i.e., material state binerization). In this study, a threshold of 0.5 was used to convert the gray-scale microstructures into binary microstructures. As mentioned earlier, in order for the (binarized) OPV microstructures to have efficient charge transport, the donor and the acceptor regions should be directly connected to the corresponding electrodes positioned at top and bottom surfaces of the thin films, respectively. In other words, the donor/acceptor pixels connected/unconnected to their respective electrodes are expected to play very different roles in the performance of the OPVs. Therefore, it was decided to define four different material states for labelling the individual pixels in the microstructures: (i) $D^{\Lambda}$ - donor pixels connected to the top surface, (ii) $D^{\circ}$ - donor pixels unconnected to the top surface, (iii) $A^{\vee}$ - acceptor pixels connected to the bottom surface, and (iv) $A^{\circ}$ - acceptor pixels unconnected to the bottom surface.  

In addition, the different types of the donor-acceptor interfaces present in the microstructure affect the charge transport in very different ways. As the charge transport occurs mainly in the connected regions, any interface between two connected regions, $I_1 = (D^{\Lambda}, A^{\vee})$, is most effective. It can also be seen that any interface between two unconnected regions, $I_2 = (D^{\circ},A^{\circ})$, is least effective. The other two types of interfaces, $I_3 = (D^{\Lambda},A^{\circ})$ and $I_4 = (D^{\circ},A^{\vee})$, are considered semi-effective.

As a final consideration, the charges created in OPV microstructures typically move through the donor and acceptor regions that are directly connected to the top and bottom electrodes ($D^{\Lambda}$ and $A^{\vee}$), respectively. In addition, if unconnected donor/acceptor regions ($D^{\circ}$ and $A^{\circ}$) are considerably close to their respective electrodes, they also play a role in the charge transport \citep{Wodo2012}. This consideration is especially important for microstructures that comprised only unconnected donor/acceptor regions. The charge transport of such microstructures is inversely related to the distance of the closest $D^{\circ}$ and $A^{\circ}$ from their relevant electrodes. 

\begin{figure}
    \centering
    \begin{subfigure}[b]{0.2\textwidth}
        \centering
        \includegraphics[width = \textwidth]{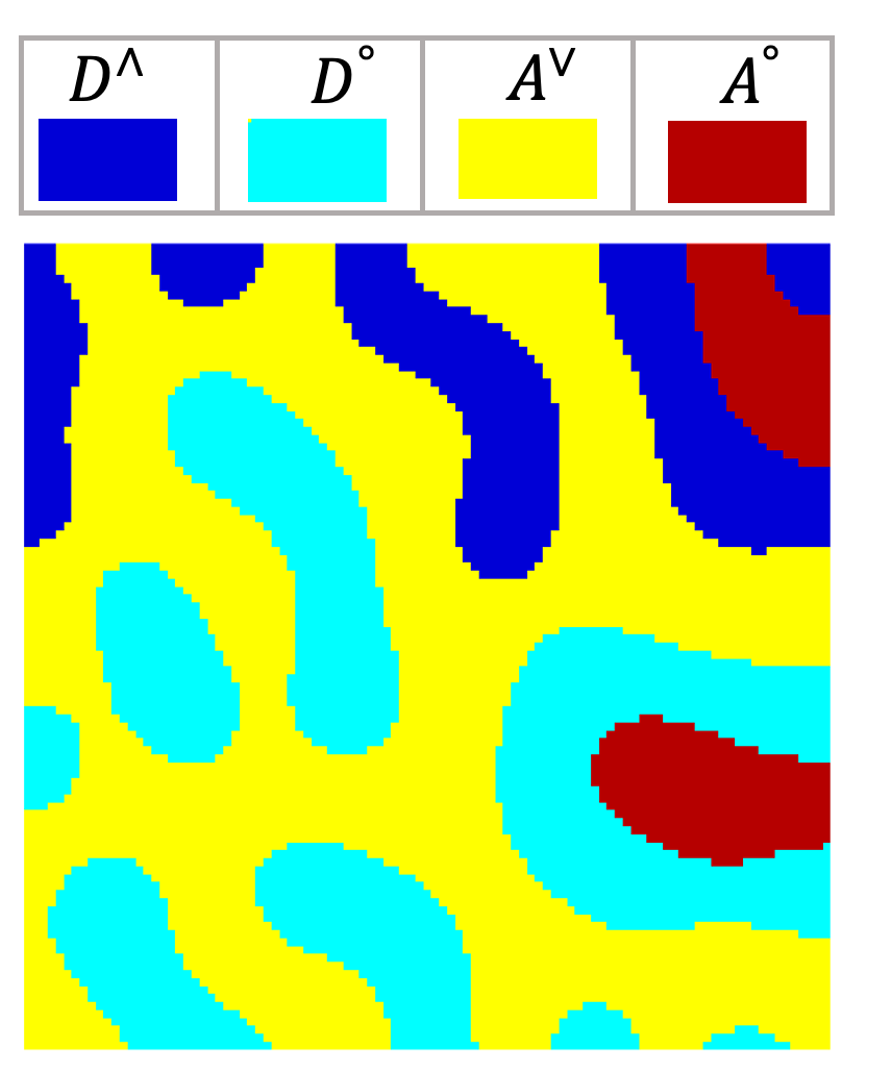}
        \caption{}
        \label{fig:4_phases}    
    \end{subfigure}
    \hspace{0.12in}
    \begin{subfigure}[b]{0.3\textwidth}
        \centering
        \includegraphics[width = \textwidth]{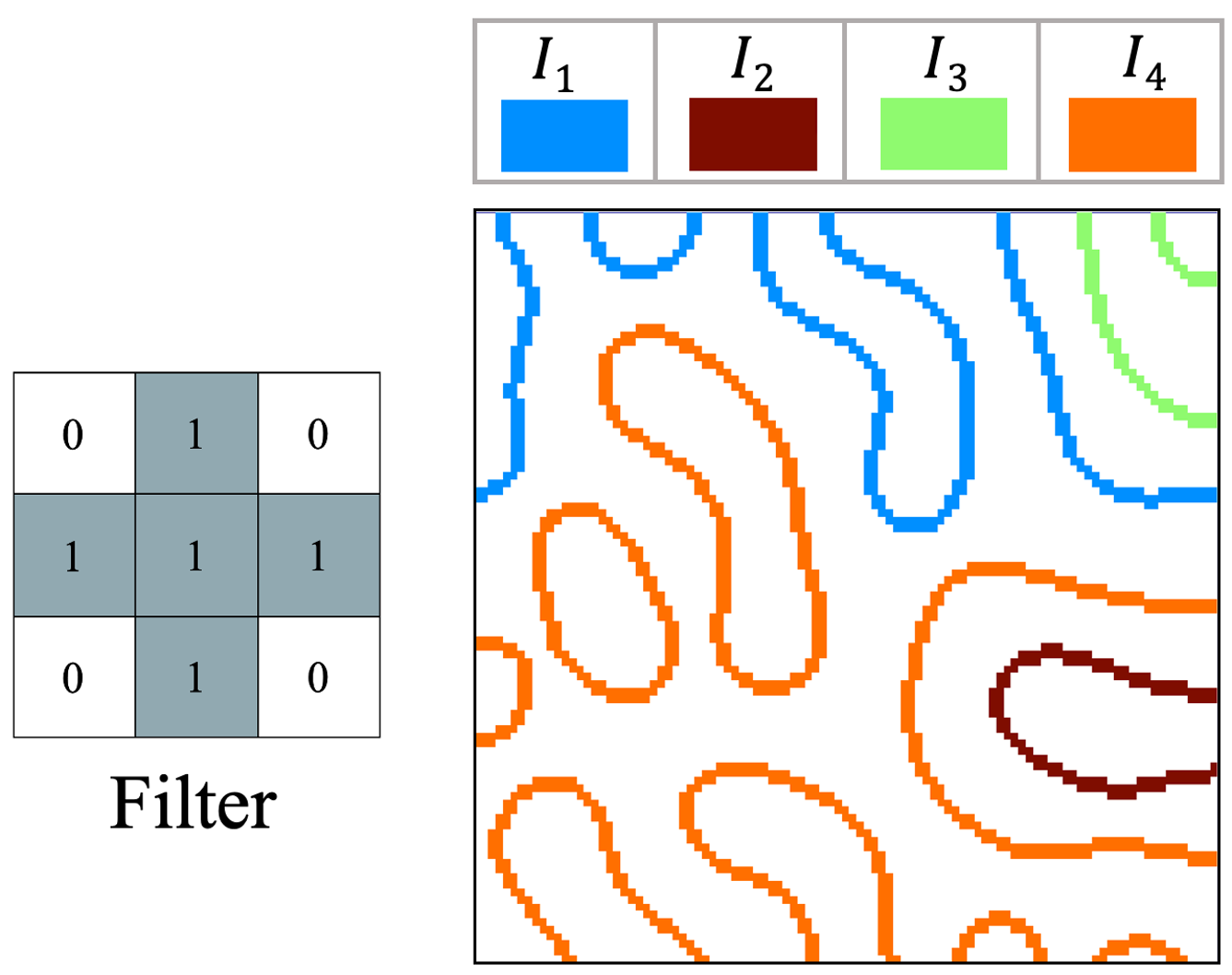}
        \caption{}
        \label{fig:4_interfaces}    
    \end{subfigure}
    \hspace{0.12in}
    \begin{subfigure}[b]{0.33\textwidth}
        \centering
        \includegraphics[width = \textwidth]{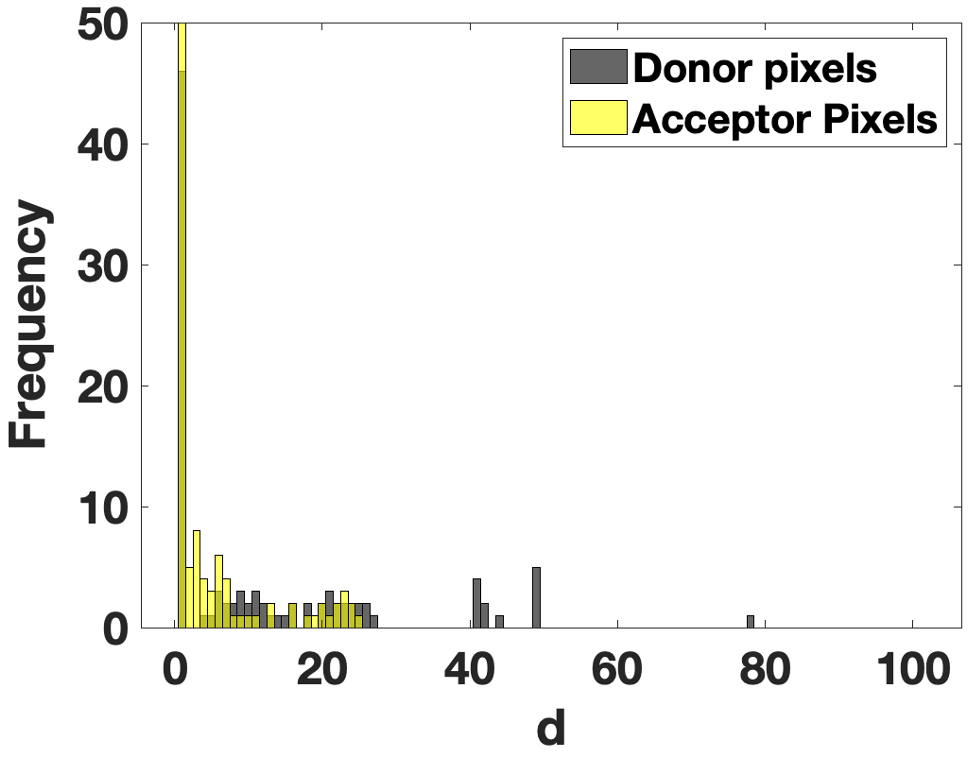}
        \caption{}
        \label{fig:2_distances}    
    \end{subfigure}
    \caption{Labelling of the material local states to each pixel of a selected OPV microstructure. (a) Each pixel is assigned one of the four material local states corresponding to connected/unconnected donor/acceptor pixels. Connectivity in this context refers to whether the donor/acceptor pixels are connected to their corresponding electrodes at the top/bottom surfaces. (b) Each interface pixel is assigned one of the four interfaces. The interfacial region is considered to be 2-pixel thick, comprising both pixels on either side of the interface. The convolution kernel used to identify the interfaces is shown on the right.  (c) A third material state is assigned to the top and bottom rows of pixels based on the shortest distance, $d$, of the donor (acceptor) pixels from the top (bottom) surface. The plot shows distribution of $d$ for the selected microstructure.}
    \label{fig:feature_id}
\end{figure}

Leveraging the insights described above, we devised and implemented a 3-step procedure to assign material local states to each pixel in each OPV microstructure.
In the first step, we assign one of the four material local states described above to each voxel in the OPV microstructure: $D^{\Lambda}$, $D^{\circ}$, $A^{\vee}$, and $A^{\circ}$ (see Figure \ref{fig:4_phases}). This was achieved by first considering the donor phase as the foreground (i.e., assigning values of one to donor pixels and zero to acceptor pixels) and  using a cluster labeling algorithm \citep{Haralock1991} to identify uniquely the connected sets of the donor pixels (i.e., donor clusters). The pixels in the donor clusters connected to the top surface were all assigned the material state $D^{\Lambda}$, while the rest of the donor pixels were assigned the material state $D^{\circ}$. A similar procedure was performed to assign the material states $A^{\vee}$ and $A^{\circ}$. Note that the assignment of these four material states is mutually exclusive. In other words, every pixel in the microstructure is assigned only one of the four material states mentioned above.

In the second step, we have defined an additional material local state identifying the different types of interfaces between the donor and acceptor pixels. This additional material state is assigned only to the interface pixels. As already described, a total of four different  interfaces are possible: ($D^{\Lambda}$, $A^v$), ($D^{\Lambda}$, $A^{\circ}$), ($D^{\circ}$, $A^v$), and ($D^{\circ}$, $A^{\circ}$) (see the microstructure shown in Figure \ref{fig:4_interfaces}). In this work, we adopted a 2-pixel interfacial region that included the first pixel on either side of the interface. The interface pixels are identified using a computational strategy developed in prior work for 2-phase microstructures \citep{Yabansu2020}. This computation is performed using the convolution kernel shown in Figure \ref{fig:4_interfaces} on selected foreground material states, which produces an integer $c_i$ at each pixel $i$. The values of $c_i \in [1:4]$ identify interfacial pixels (i.e., the interior pixels within the foreground and background would have values zero and five, respectively). In this work, special considerations were made to account for the non-periodicity of the microstructures. Specifically, this challenge was addressed using suitable zero-padding schemes \citep{Cecen2016}. For non-periodic microstructures, the sets of edge pixels and corner pixels were identified separately; edge pixels with $c_i \in [1:3]$ and corner pixels with $c_i \in [1,2]$ denote interfacial pixels. By applying the procedure described above to each phase (i.e., treating each phase as foreground one at a time), each interface pixel can be mapped uniquely to one the aforementioned four types of interfaces.  Figure \ref{fig:4_interfaces} shows the labelling of the interface pixels for the example microstructure shown in Figure \ref{fig:4_phases}.

In the last step of the unsupervised feature identification procedure employed in this study, we identify a third material state descriptor for the top/bottom rows of pixels connecting to the electrodes. This feature is designed to capture the effect arising from the shortest distance of donor/acceptor pixels from their respective electrodes. As already mentioned, this feature is especially important for microstructures where the donor/acceptor pixels are not in direct contact with their corresponding electrodes. Figure {\ref{fig:2_distances}} presents a histogram of the shortest vertical distance of the donor (acceptor) pixel to the top (bottom) surface, $d$, for the example microstructure shown in Figure \ref{fig:4_phases}. It was decided to use  $\exp({-d/\lambda})$ as the feature value for each top/bottom pixel, with $\lambda=10$ nm reflecting the expected diffusion length for exciton transport \citep{shaw2008exciton,Wodo2012}. Consequently, the feature value is one when the pixels are in direct contact and exponentially decreases when there is a gap.

After labelling the material local states, the next step involves the computation of the important microstructure statistics. The central challenge comes from the large number of spatial statistics that could be computed. In the present case, since there are a total of eight material local states (four acceptor/donor states and four interface states), one can potentially define a total of $8^2 = 64$ sets of spatial correlations (including auto-correlations and cross-correlations). Since each set of spatial correlations has a total of $101 \times 101 = 10,201$ features, the full set of features becomes unwieldy for establishing surrogate models. In prior work \citep{Yabansu2020} on correlating the effective permeability of a porous solid to its pore structure, it was observed that the auto-correlations of the material local states (including interface states) were adequate for producing high fidelity structure-property linkages. Utilizing the insights from that work, we have included only the following sets of spatial correlations in establishing the surrogate models presented in this work: i) 2-point spatial auto-correlations for each of the four main material local states $\big\{f_{\boldsymbol{r}}^{D^{\Lambda} D^{\Lambda}}$, $f_{\boldsymbol{r}}^{D^{\circ} D^{\circ}}$,$f_{\boldsymbol{r}}^{A^{\vee} A^{\vee}}$, $f_{\boldsymbol{r}}^{A^{\circ} A^{\circ}}\big\}$, and ii) 2-point spatial auto-correlations for each of the four interfacial local states $\big \{f_{\boldsymbol{r}}^{I_1 I_1}$, $f_{\boldsymbol{r}}^{I_2 I_2}$, $f_{\boldsymbol{r}}^{I_3 I_3}$, $f_{\boldsymbol{r}}^{I_4 I_4}\big \}$. Even using only this subset of spatial correlations produces a total of $8 \times 10,201 = 81,608$ features. As already described in Section \ref{sec:MKS_featureENG}, PCA is applied to obtain a small number of features (i.e., PC scores) as inputs to the surrogate structure-property models. Prior to application of PCA, each of the eight sets of spatial correlations are scaled to exhibit the same variance across the entire dataset. This is necessary due to the fact that PCA aims to capture the variance in the dataset in the smallest number of terms. Therefore, scaling the different sets of spatial correlations ensures that each set of spatial correlations is equally weighted in the PC representations. In this work, for reasons already explained, the averaged values of $\exp({-d/10})$ for both electrodes, denoted as $\big \{{\delta}^{\Lambda}$,  ${\delta}^{\vee}\big \}$, are used as additional features (i.e., these are appended to the selected PC scores representing the microstructure statistics as additional features).
    
\subsection{Local Gaussian process surrogate models for OPVs}

The microstructure PC scores as well as the two distance-based features are used as inputs to train a local Gaussian process (laGP) surrogate model to predict the short circuit current of OPV microstructures. Each input is scaled to exhibit the same variance across the entire ensemble of the dataset. This is needed because laGP models identify local subsets of the training data using suitable distance measures. For each target point, the first $n_0$ closest neighboring points are chosen as the initial training set for building the initial GP. Subsequently, the ALC criterion is used to sequentially update the training data to maximize the expected information gain. As the training subset is sequentially updated, one expects to see a systematic decrease in the improvement to the model performance. Consequently, one would naturally reach a point where further updating the training set would only minimally improve the laGP model performance. In this study, the sequential update of the laGP model was continued until the reduction in the prediction variance was smaller than $10^{-6}$. In the protocol described above, the final size of the local training set is denoted as $n_d=n_0+n_{ALC}$, where $n_{ALC}$ denotes the number of training points selected using the ALC criterion. The performance of the trained laGP models produced in this work was quantified using multiple error measures, including normalized mean absolute error ($nMAE$), normalized median absolute deviation ($nMAD$) and $R^2$. These are defined as

\begin{align}
    &nMAE = \frac{\frac{1}{N} \sum_{i=1}^{N} |J_{sc}^{(i)} - \tilde{J}_{sc}^{(i)}|}{\bar{J}_{sc}}\\
    &nMAD = \frac{\text{median} \bigg( |J_{sc}^{(1)} - \tilde{J}_{sc}^{(1)}|,|J_{sc}^{(2)} - \tilde{J}_{sc}^{(2)}|,\cdots,|J_{sc}^{(N)} - \tilde{J}_{sc}^{(N)}| \bigg)}{\bar{J}_{sc}} \\
    &R^2 = 1 - \frac{\sum_{i=1}^{N} (J_{sc}^{(i)} - \tilde{J}_{sc}^{(i)})^2}{\sum_{i=1}^{N} (J_{sc}^{(i)} - \Bar{J}_{sc})^2}
\end{align}

\noindent where  $J_{sc}^{(i)}$ and $\tilde{J}_{sc}^{(i)}$ are the actual (ground truth) and the predicted short circuit current of the $i^{\text{th}}$ target point, and  $N$ is the number of test points. $\Bar{J}_{sc}$ denotes the mean value of the $J_{sc}$ values. $R^2$ serves as an indicator of how much of the variation in the output is explained by the inputs. The value of $R^2$ for a perfect model is expected to be one. Likewise, for a flat line model that always predicts the mean, the value of $R^2$ will be zero.

\section{Results and discussions}
\label{sec:Res}

In the present study, an ensemble of 33,552 distinct OPV microstructures was generated to establish the desired data-driven microstructure-property linkage for OPV films. The short circuit current $J_{sc}$ associated with each microstructure was obtained by solving the XXD equations discussed in Section \ref{sec:data_gen}. The unsupervised feature engineering framework described in Section \ref{sec:feature_eng} was employed on each microstructure.

\begin{figure}
    \centering
    \includegraphics[width=1\textwidth]{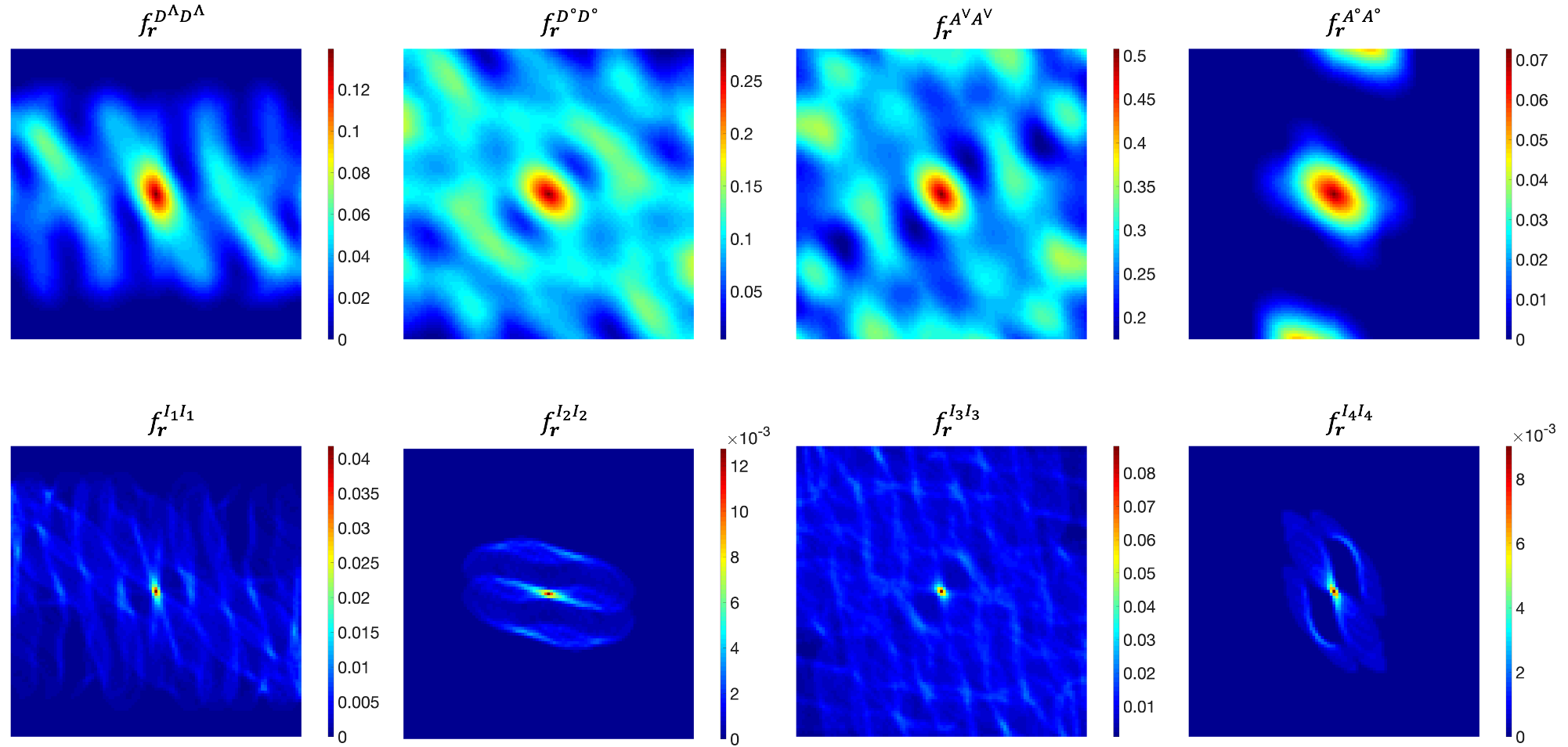}
    \caption{The 8 sets of 2-point spatial auto-correlations corresponding to main material states and interfaces of the example microstructure shown in Figure \ref{fig:feature_id} is shown. The center value of these statistical maps is volume fraction of the corresponding material state.}
    \label{fig:2_pts}
\end{figure}

Figure \ref{fig:2_pts} depicts the eight sets of spatial auto-correlations computed for the example microstructure shown in Figure \ref{fig:4_phases}. The top and bottom rows in this figure present spatial auto-correlations of the four main material states and the four interface states, respectively. Note that the auto-correlations exhibit centro-symmetry, because the values of the statistics for $\boldsymbol{r}$ and $-\boldsymbol{r}$ are the same. Therefore, half the information in these maps is redundant and could be eliminated before performing the PCA.  The central peak value in each  auto-correlation map, corresponding to $\boldsymbol{r} = \boldsymbol{0}$, reflects the volume fraction of the specific material state. For the interface states, this value corresponds to the volume fraction occupied by the 2-voxel wide interface regions defined in this work. The auto-correlation maps implicitly capture a significant amount of statistical information on the shape, size, and spacing distributions of the material states in the microstructure. For instance, the bands in the $f^{D^{\Lambda}D^{\Lambda}}_{\boldsymbol{r}}$ map capture important features related to the size, shape, orientation, and spacing of the $D^{\Lambda}$ regions in the microstructure (compare the auto-correlation map with the actual microstructure in Figure \ref{fig:4_phases}). Similarly, $f^{A^{\circ}A^{\circ}}_{\boldsymbol{r}}$ captures the details of the more compact and isolated positioning of the $A^{\circ}$ regions in this microstructure. In contrast, the auto-correlation maps for $D^{\circ}$ and $A^{\vee}$ indicate that these regions are more broadly distributed in the microstructure.  Similar observations can be made for the auto-correlation maps of the interface states.

\begin{figure}
    \centering
    \includegraphics[width=1\textwidth]{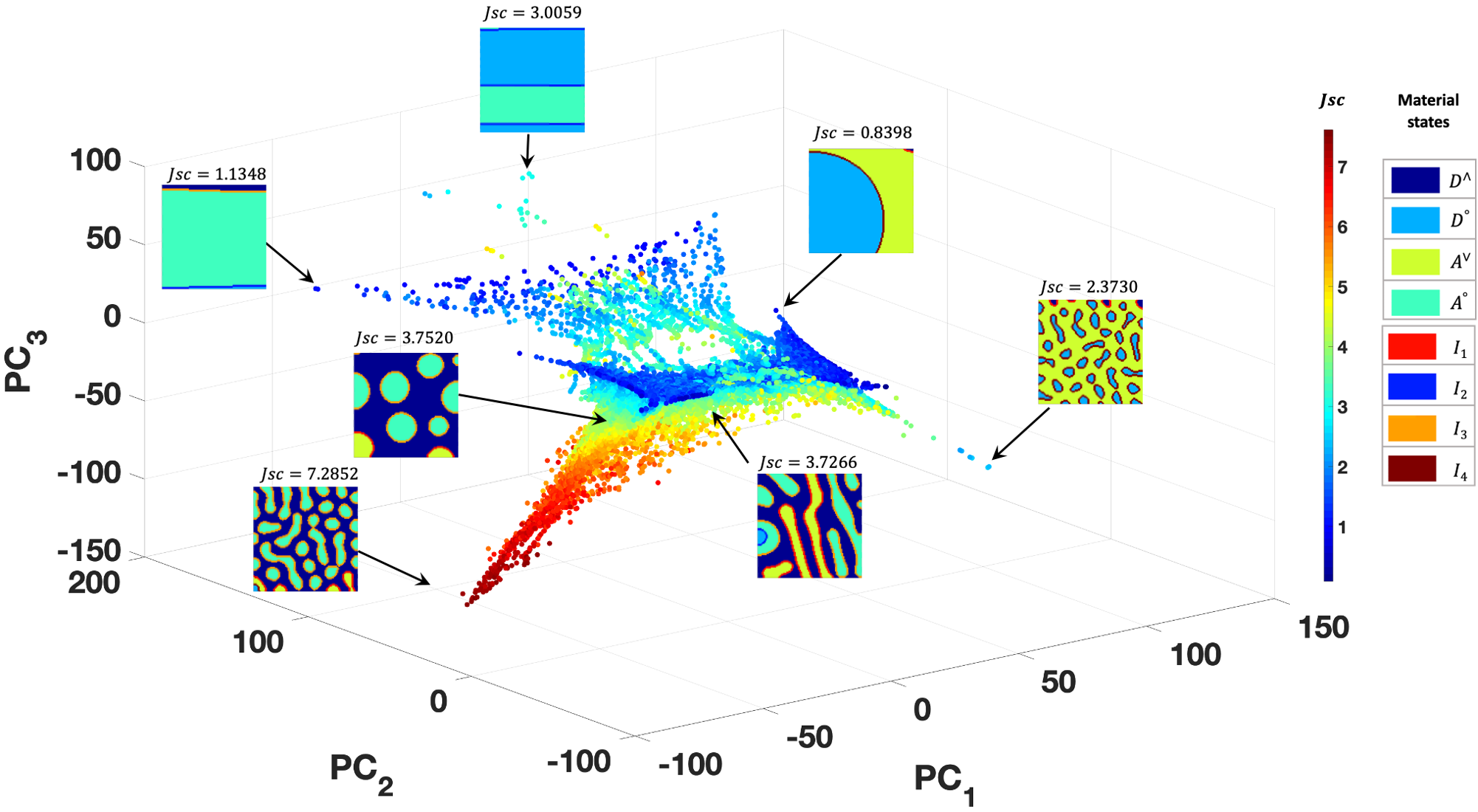}
    \caption{The low-dimensional representation of the entire data ensemble of OPV microstructures in the first 3 PC basis is depicted. The PC representations are truncated after the first 10 PCs. The unsupervised PCA is powerful in capturing the microstructural differences as well as the variance in the values of $J_{sc}$. }
    \label{fig:PCA}
\end{figure}

In order to efficiently compute the PCA of the large data matrix of size $33,552 \times 81,608$ assembled in this work, we took advantage of the randomized SVD algorithm implemented in DASK package in Python programming language \citep{Dask2015}. It was decided to truncate the PC representations obtained from this protocol to 10 PCs, because there was no appreciable improvement in the variance captured beyond this truncation level. This represents a significant reduction in the dimensionality of the microstructure representation, where we started with 81,608 spatial correlations and ended up with only 10 PC scores. The representation of all 33,552 microstructures in the first three PCs is presented in Figure \ref{fig:PCA}. In this figure, each data point corresponds to the first three PC scores of the microstructure statistics and is colored using its value of $J_{sc}$. Although the three PC scores represent only a subset of the regressors we intend to use in this work (a total of ten PC scores and two distance-based metrics will be used), it is very encouraging to see the patterns in Figure \ref{fig:PCA} suggesting a strong dependence of the target on these regressors. 

A direct interpretation of the PC scores is currently not possible. Essentially, each PC basis represents a linearly weighted collection of 81,608 spatial correlations. The PC score of each OPV microstructure represents the projection (i.e., dot product) of its set of 81,608 spatial correlations on the corresponding PC basis. The high dimensionality of the PC basis makes it impractical to seek the precise physical meaning of the PC scores. However, it was found that the first PC score is highly correlated to the volume fractions of the four main material local states as well as the $I_1$ and $I_3$ interface states (interfaces of $D^{\Lambda}$ with acceptor material states). The second PC score was found to be highly correlated to volume fractions of $I_2$ and $I_4$ interface states (interfaces of $A^{\vee}$ with donor material states). In addition to the information on the volume fractions, PC scores contain rich information on other morphological aspects of microstructures such as shape, size, orientation and spacing of material features within OPV microstructures. For instance, as seen from Figure \ref{fig:PCA}, the microstructures comprising coarser regions of $A^{\vee}$ and/or $D^{\circ}$ have higher $PC_1$ values, while the microstructures with coarser regions of $A^{\circ}$ and/or $D^{\Lambda}$ have smaller $PC_1$ values. Several other similar qualitative observations can be made by inspecting Figure 4 closely. As another example, it can been seen that microstructures comprised mainly/only from coarser unconnected donor/acceptor regions are completely separated from the microstructures with connected donor/acceptor finer regions in the low-dimensional PC representation.

\begin{figure}
    \centering
    \begin{subfigure}[b]{0.575\textwidth}
        \centering
        \includegraphics[width = \textwidth]{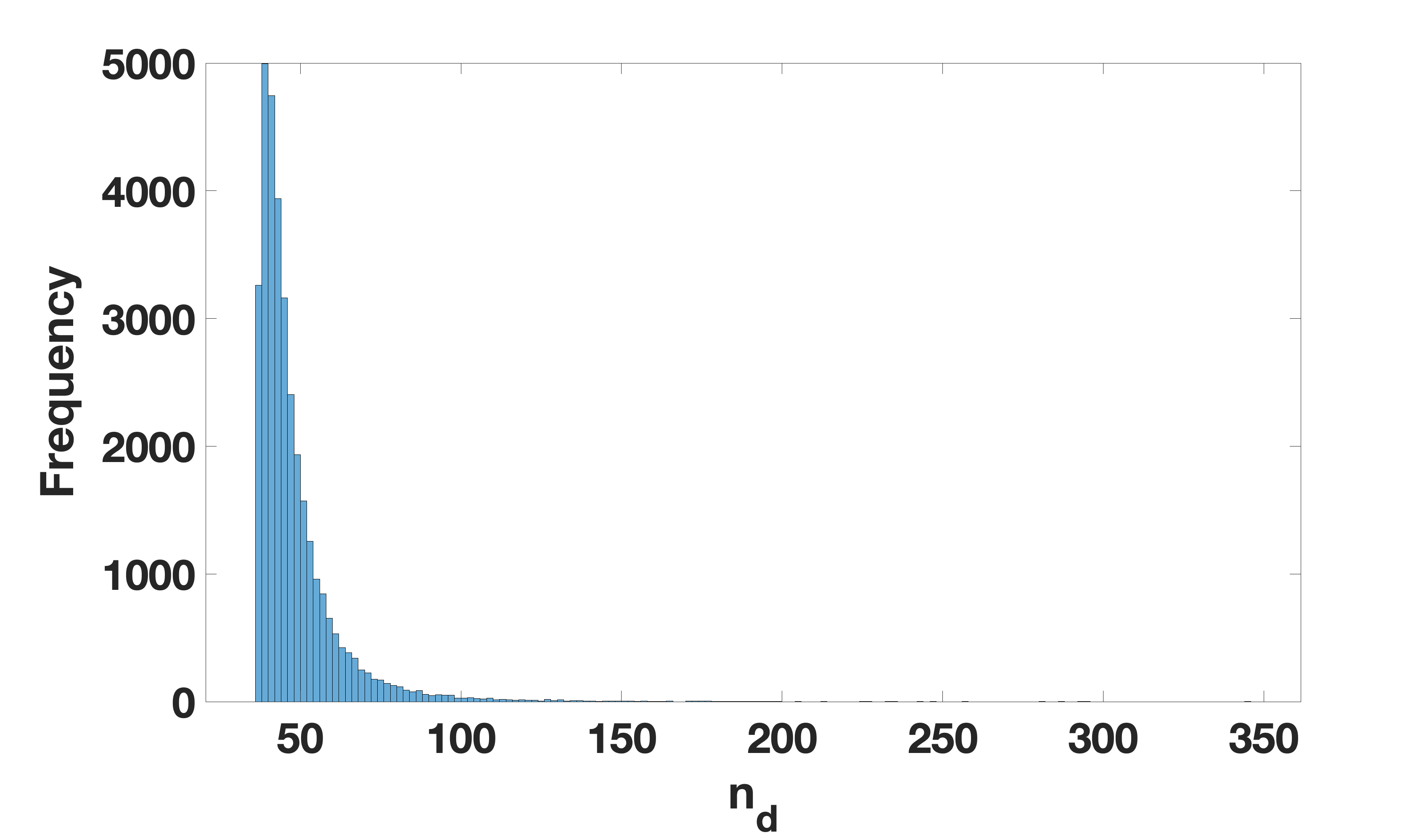}
        \caption{}
        \label{fig:Dsize}
    \end{subfigure}
    \begin{subfigure}[b]{0.9\textwidth}
        \centering
        \includegraphics[width = \textwidth]{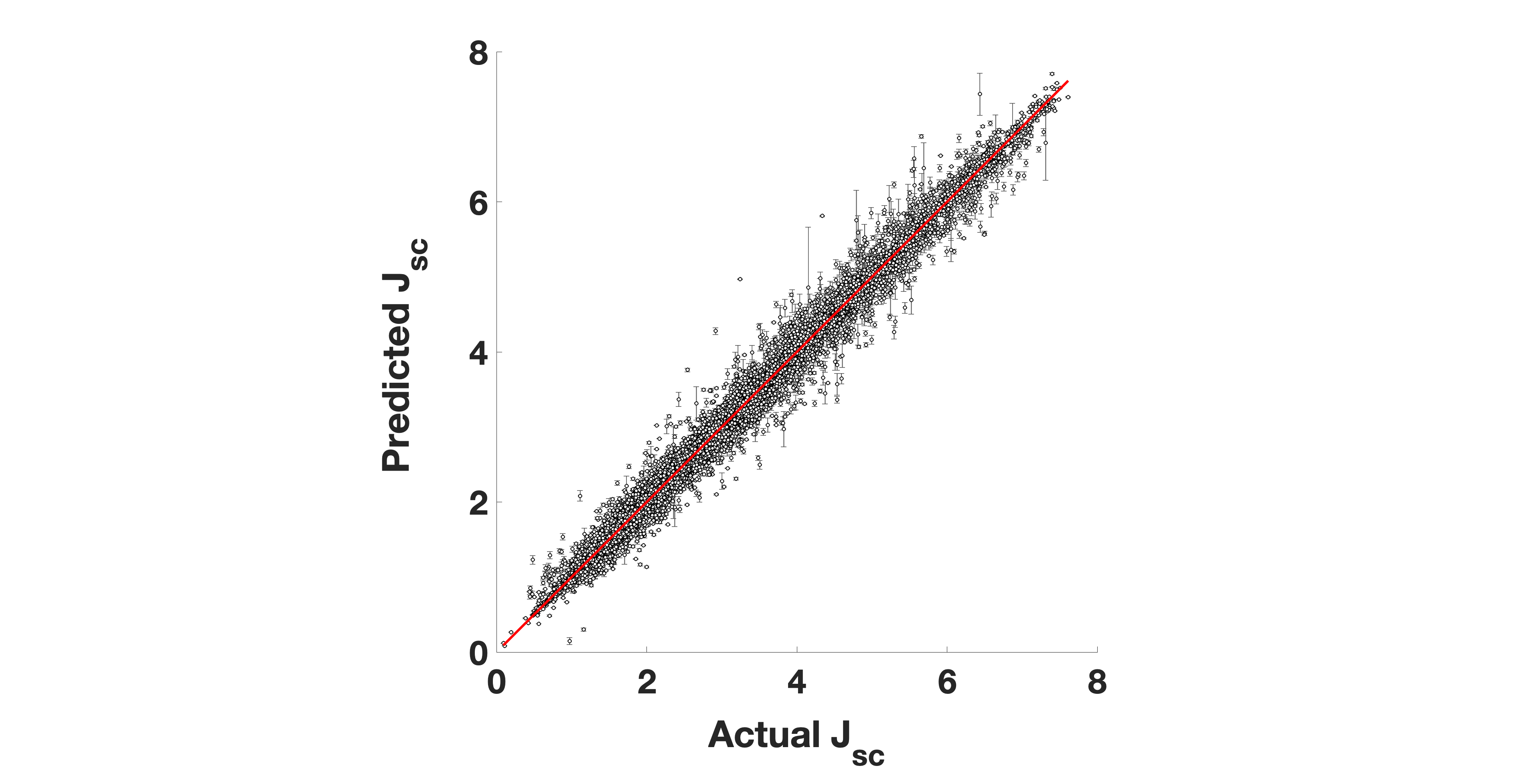}
        \caption{}
        \label{fig:parity}    
    \end{subfigure}
    \begin{subfigure}[b]{0.65\textwidth}
        \centering
        \includegraphics[width = \textwidth]{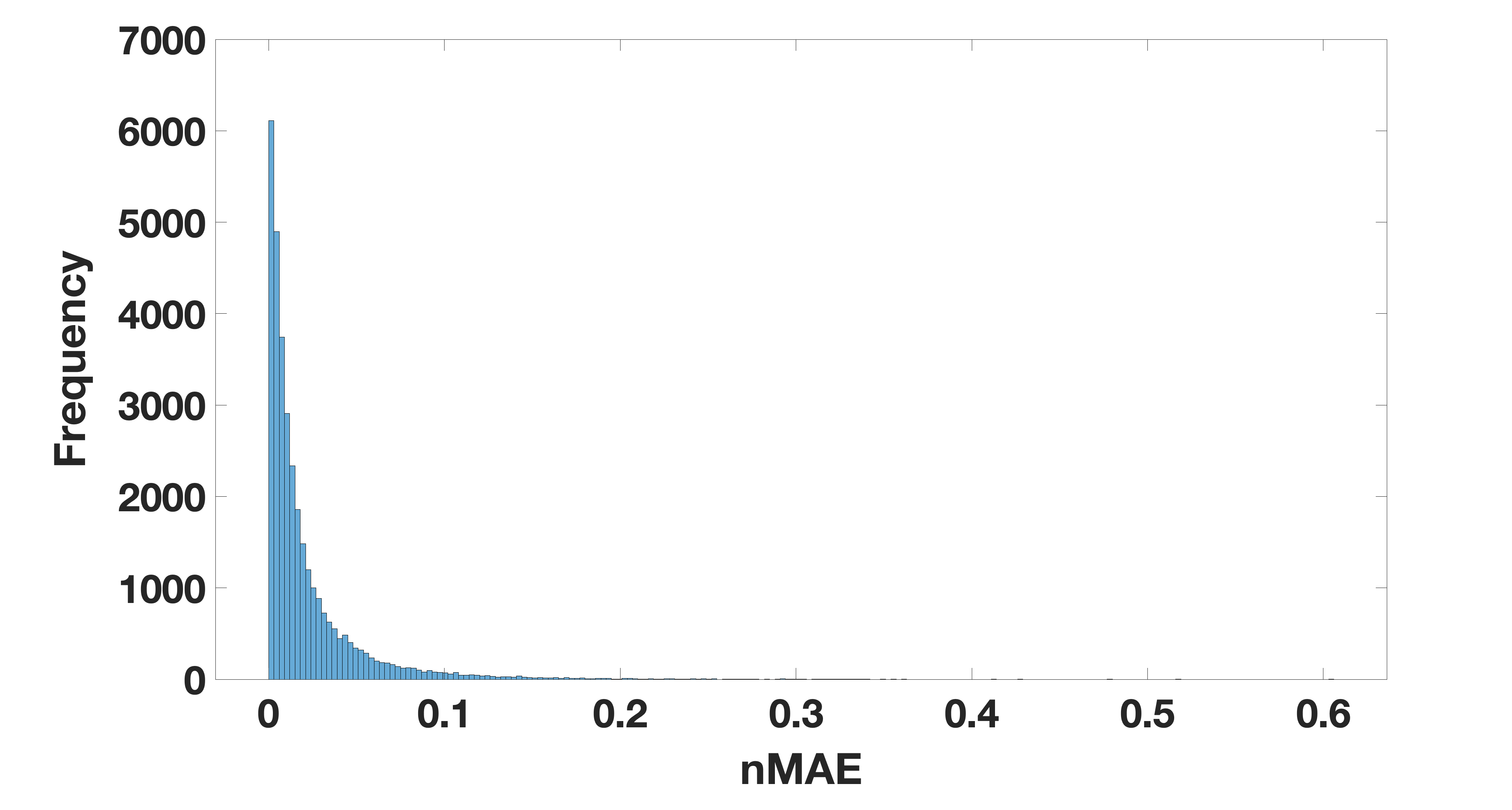}
        \caption{}
        \label{fig:AE}    
    \end{subfigure}
    \caption{Depiction of the performance of the data-driven structure-property linkages trained for OPV microstructures is presented. The total design size of each laGP model is determined using ALC criterion. The distribution of the final design size is shown in (a). The parity plot comparing the predictions and actual values of $J_{sc}$ as well as the relative mean absolute error are presented in (b) and (c), respectively.The established high-fidelity microstructure-property linkages demonstrate the utility of the developed feature engineering framework for organic photovoltaics.}
    \label{fig:S-P}
\end{figure}

The 10 microstructure PC scores and the two averaged distance-based metrics, ${\delta}^{\Lambda}$ and ${\delta}^{\vee}$, are used as inputs to train the surrogate laGP models using the R package language \citep{lagp2016}. As already noted, each input feature is scaled to exhibit the same variance across the entire dataset for this model building strategy. A laGP model is produced for each test point using a set of $n_0=35$ closest neighbors in the input domain. The ALC criterion is employed to sequentially add points to the design space such that their addition maximizes the expected information gain. The training size for the 33,552 laGP models produced in this study was in the range $[36,346]$. The  distribution of the training sizes is shown in Figure \ref{fig:Dsize}. It is seen that more than 99\% of the laGP models built in this study needed less than 100 local training data points. This small size of the local training data set significantly reduces the computational cost involved in building the desired laGP models. Figures \ref{fig:parity} and \ref{fig:AE} present the parity plot comparing the $J_{sc}$ predictions from the laGP models with their corresponding ground-truth values as well as the uncertainty associated with the model predictions (i.e., one standard deviation from the mean prediction shown as error bars) and the distribution of the relative mean absolute errors, respectively. The standard deviation in $99.7\%$ of the trained models is within $5\%$ of the $\bar{J}_{sc}$. Those few models that exhibit higher uncertainties correspond to the microstructures that fall on the boundary of the input PC domain. This is to be expected as laGP performs better in the interior of the input domain, compared to the edges of the input domain (there is limited availability of training points in these regions). The normalized absolute error was higher than 0.05 in only $8\%$ of the trained laGP models. Considering the entire set of trained models, the normalized mean absolute error $nMAE$ and the normalized mean median absolute deviation $nMAD$  were  $2.16\%$ and $1.10\%$, respectively. Moreover, a high value of $R^2=0.99$ was calculated for the trained laGP models, which demonstrates that a high proportion of the variance in the target is being captured well by the model inputs. This clearly demonstrates the efficacy of the novel feature engineering framework presented in this work in establishing high-fidelity data-driven microstructure-property mappings in organic photovoltaics.

\section{Conclusions}
\label{sec:Con}

A novel unsupervised feature engineering framework for data-driven mappings of microstructures to their photovoltaic properties has been successfully developed. A computationally  efficient labeling  of two sets of salient material states (four bulk material states and four interface states) served as critical features, and were extracted via digital image processing techniques. In order to take into account the non-periodicity of the OPV microstructures, suitable zero-padding schemes were utilized. It was found that only 2-point spatial auto-correlations of the eight sets of labeled material states were sufficient to extract reasonably accurate structure-property linkages. The low-dimensional representation of this rich large set of material statistics was then obtained from principal component analysis by taking advantage of a scalable randomized SVD algorithm. In addition to material PC scores, it was found that two additional expert-defined distance-based metrics further improved the accuracy of the data driven structure-property linkages. A localized-version of the Gaussian process (laGP) was employed to extract these data-driven reduced order structure-property linkages. The laGP model took advantage of active learning to detect a small subset of the training data used to build separate models for each target data point. It was shown that with only a small subset of the training dataset one can build accurate laGP models. This significantly reduced the computational cost involved in building of the desired laGP model. The uncertainty associated with the model predictions were quantified by considering one standard deviation from the mean prediction. It was found that the uncertainty of only $0.3\%$ of the trained models is higher than $5\%$ of the mean value of the target property. The high-fidelity accurate structure-property linkages extracted in this study attest to the tremendous efficacy of the proposed novel feature engineering framework for complex organic photovoltaic microstructures.

\section{Acknowledgments}

The authors acknowledge funding from NASA Langley Research Center. BG acknowledges partial support from NSF 1906194 and ONR Award N00014-19-1-2453. SK acknowledges partial support from Vannevar Bush Fellowship through ONR Award N00014-18-1-2879.

\bibliographystyle{elsarticle-num-names}
\biboptions{sort&compress} 
\bibliography{cas-refs}





\end{document}